\documentclass[aps,prd,twocolumn,floatfix,noshowpacs,tightenlines,noshowkeys,superscriptaddress,amsmath,amssymb,nofootinbib]{revtex4-1}
\usepackage{amssymb,amsbsy,epsfig,color,graphicx}
\usepackage{color}
\usepackage{[longtable}
\usepackage{array}
\usepackage{dcolumn}   
\usepackage{cellspace}
\usepackage{mathtools}
\usepackage{amstext}
\usepackage{amssymb}
\usepackage{stmaryrd}
\usepackage{stackrel}
\usepackage{graphicx}
\usepackage{esint}
\usepackage[utf8]{inputenc}
\usepackage{adjustbox}
\usepackage{multirow}
\usepackage{float}
\usepackage{booktabs}
\usepackage{enumitem}
\usepackage{natbib}
\usepackage{slashed}
\usepackage{textgreek}
\usepackage[hidelinks]{hyperref}

\hypersetup{
  colorlinks   = true, 
  urlcolor     = blue, 
  linkcolor    = blue, 
  citecolor   = red 
}
\usepackage{etoolbox} 
\usepackage{lipsum} 
\usepackage[capitalize]{cleveref}
\usepackage{lipsum,booktabs}
\usepackage{changepage}
\usepackage{multirow}
\usepackage[caption=false]{subfig}
\allowdisplaybreaks
\usepackage{soul}
\usepackage{ragged2e}
\usepackage{bm}
\begin{document}

\preprint{APS/123-QED}

\title{Propagation of Enzyme-driven Active Fluctuations in Crowded Milieu}


\author{Rik Chakraborty}
\affiliation{Laboratory of Soft and Living Materials, Department of Physics, Indian Institute of Technology Gandhinagar, Palaj, Gujarat 382055, India}

\author{Arnab Maiti}
\affiliation{Laboratory of Soft and Living Materials, Department of Physics, Indian Institute of Technology Gandhinagar, Palaj, Gujarat 382055, India}


\author{Diptangshu Paul}
\affiliation{Department of Mechanical Engineering, Indian Institute of Technology Gandhinagar, Palaj, Gujarat 382055, India}


\author{Rajnandan Borthakur}
\affiliation{Department of Mechanical Engineering, Indian Institute of Technology Gandhinagar, Palaj, Gujarat 382055, India}

\author{K. R. Jayaprakash}
\affiliation{Department of Mechanical Engineering, Indian Institute of Technology Gandhinagar, Palaj, Gujarat 382055, India}

\author{Uddipta Ghosh}
\affiliation{Department of Mechanical Engineering, Indian Institute of Technology Gandhinagar, Palaj, Gujarat 382055, India}

\author{Krishna Kanti Dey}
\email{Author to whom correspondence should be addressed:k.dey@iitgn.ac.in}
\affiliation{Laboratory of Soft and Living Materials, Department of Physics, Indian Institute of Technology Gandhinagar, Palaj, Gujarat 382055, India}
\date{\today}

\newcommand{\tend}[1]{\hbox{\oalign{$\bm{#1}$\crcr\hidewidth$\scriptscriptstyle\bm{\sim}$\hidewidth}}}
\newcommand{\tenq}[1]{\hbox{\oalign{$\bm{#1}$\crcr\hidewidth$\scriptscriptstyle\bm{\approx}$\hidewidth}}}
             
\begin{abstract}

We investigated the energy transfer from active enzymes to their surroundings in crowded environments by measuring the diffusion of passive microscopic tracers in active solutions of ficoll and glycerol. Despite observing lower rates of substrate turnover and relatively smaller enhancement of passive tracer diffusion in artificial crowded media compared to those in aqueous solutions, we found a significantly higher relative diffusion enhancement in crowded environments in the presence of enzymatic activity. Our experimental observations, coupled with supporting analytical estimations, underscored the critical role of the intervening media in facilitating mechanical energy distribution around active enzymes.
\end{abstract}

\maketitle
\section{\label{sec:level1}INTRODUCTION}\protect\
The ability of active enzymes to generate mechanical fluctuations during catalytic turnovers has been recently a focus area of investigation in active matter physics.~\cite{1,2,3,4,5,6} Several experimental~\cite{7,8,9,10} and theoretical~\cite{11,12,13,14} studies, have demonstrated that active enzymes, while catalysing substrates, could enhance their diffusion under aqueous environment, the exact mechanism of which remains still to be fully understood. The observations however, opened a new area of investigation in molecular biophysics, where mechanical energy transduction by single molecule was previously believed to be unlikely.~\cite{15} While the enhancement of catalytic enzymes' diffusion was subject to both positive and negative arguments~\cite{16,17,18,19,20,21}, it was theoretically proposed~\cite{22,23,24,25,26,27} and experimentally demonstrated~\cite{28,29,30,31} that active fluctuations generated by enzymes during catalysis could also influence the dynamics of nearby passive tracers through long-ranged interactions. The observed capability of active enzyme molecules to transduce mechanical energy from catalytic turnover and thereby influence the dynamics of their surroundings, akin to the ATP-driven molecular motors, presents an intriguing avenue for further exploration. The observations were important as molecular motors have been reported to drive diffusive-like, non-thermal motion of cellular components, influencing the overall metabolic state of the cell.~\cite{32,33} The dynamic coupling observed between catalytic enzymes and their surroundings provides compelling evidence that these molecules, beyond their primary role as catalysts in biochemical reactions, have secondary functions and may serve as mediators of dynamic interaction pathways.~\cite{34} However, considering that the cytoskeletal transport by motor proteins occurs in highly dissipative and crowded environments,~\cite{35,36,37} it is imperative to investigate whether the mechanical forces generated by the enzymes are perceivable in dense and crowded intracellular environments. A positive answer to this question would also warrant investigation to check the influence of the crowded environment on the generation and propagation of reaction-induced fluctuations, and assess the potential implications of such active processes on intracellular particle dynamics. Successful exploration of such phenomena is expected to uncover new research avenues, significantly advancing our understanding of activity-induced transport, assembly, and other related phenomena. With this broader perspective in mind, in this study, we studied the mechanical energy transfer from active enzymes within a dense and crowded environment by estimating the diffusion of microscopic passive tracers suspended in such media. To mimic intracellular viscosity and crowding, we used molecular crowders such as ficoll 400 and glycerol, as commonly used in other studies.~\cite{38,39} The diffusion of passive tracers was considerably lower in the crowder solutions compared to that in water, primarily due to the higher viscosity of the former. Notably, enzymatic substrate turnover rates in crowded environments were also slower than those in aqueous media, which could further contribute to reducing passive tracer diffusion in these media compared to that in water. Interestingly, upon initiating enzymatic reactions, we observed enhanced diffusion of passive tracers, even in crowded environments, similar to observations in aqueous solutions.~\cite{28,30} However, the relative increase in diffusion in crowded environments, observed with both ficoll and glycerol, was significantly greater than that in aqueous media. We conducted experiments measuring tracer diffusion in three different media (water, ficoll, and glycerol) under different reaction rate conditions. Based on the results obtained in these experiments we hypothesized that artificial crowded environments facilitate a greater transfer of reaction-generated fluctuations around active enzymes compared to dilute aqueous environments.

In order to validate our hypothesis, we developed a model considering that the binding and unbinding of substrates in enzymes largely governed their conformational fluctuations resulting in the generation of stochastic forces. The actual conversion of reactant to products, on the other hand, was likely to contribute to the net thermal energy of the system, which was previously found to be insignificant in enhancing single enzyme diffusion.~\cite{40} The conformation change-induced forces propagated through the surrounding medium in the form of high-frequency pressure pulses. We assumed the active fluctuations propagating around enzymes' vicinity as spherical pressure waves and computed the corresponding average energy density at specific distances away from the point of actuation. The analytical estimates found nice qualitative match with the experimental results, suggesting that the intervening media played a crucial role in deciding the energy propagation around catalytically active enzymes, which might have important implications over various biochemical processes in active intracellular milieu.

\section{\label{sec:level2}RESULTS AND DISCUSSIONS}\protect\
3~$\mu$m sized polystyrene latex beads (LB30, 10\% solids), catalase from bovine liver (catalogue no: C9322), urease from Canavalia ensiformis (Jack bean) (catalogue no: U1500), urea (catalogue no: U5128), phenol red (catalogue no: P3532), glycerol (catalogue no: G5516), ficoll PM 400 (catalogue no: F4375) were purchased from Sigma-Aldrich. 30\% (w\slash v) hydrogen peroxide (H$_2$O$_2$) solution (catalogue no: 107209) and 95-97\% (w/w) sulphuric acid (H$_2$SO$_4$) solution (catalogue no: 100731) were purchased from Merck.
To prepare the experimental tracer solution, $10~\mu \mathrm{L}$ of the stock solution was diluted in a series of steps, with each step having a dilution ratio of $1:10$, using deionized (DI) water. This resulted in a secondary stock solution with a solid concentration of 0.1\%. The secondary stock solution was then further diluted using DI water (pH~7 and resistivity of 18.2~$\mathrm{M}\Omega \cdot \mathrm{cm}$) or 10\% (v\slash v) glycerol or 3\% (w\slash v) ficoll 400 to prepare the final experimental solutions containing 0.005\% polymer particles. For experiments, all measurements were conducted at room temperature of $25~^{\circ}$C.
Commercially procured glass slides (Blue Star, PIC-1 microslides, $76 ~\mathrm{mm}~\mathrm{L} \times 26~ \mathrm{mm}~\mathrm{W} \times 1.35~\mathrm{mm}~\mathrm{H}$), were washed thoroughly using Labolene solution, followed by sonicating them for 10 min. After sonication, the glass slides were treated with piranha solution, which is a mixture of H$_2$SO$_4$ and H$_2$O$_2$ in the ratio of $7:3$, for a period of 1 h. Next, the slides were removed carefully from the piranha solution and washed with deionized (DI) water. Finally, they were dried using an air drier. A small circular plastic chamber was then securely attached to the glass slide using suitable adhesive followed by air drying at room temperature.

We used optical video microscopy and single particle tracking technique to characterize the motion of polymer particles over the entire field of view in a quasi-2D condition. Experimental videos were recorded using a Nikon Eclipse Ti2 Inverted Microscope, with an objective of 60$\times$, 10 frames per second. Each video was recorded for 5 min after allowing a 30 min stabilization period to ensure that it had reached a steady state. For active samples, after the stabilization period, an additional 5 min were given to minimize the effects of perturbations arising due to substrate addition. Frames from each recorded video were analyzed and tracked through particle tracking codes developed by Crocker and Grier.~\cite{41} Mean squared displacements (MSD) and diffusion coefficients were estimated using a code developed in MATLAB. The results obtained at different experimental conditions were plotted and fitted using OriginPro.

The dynamic viscosities of DI water, 10\% (v\slash v) glycerol and 3\% (w\slash v) ficoll 400 at $25~^{\circ}$C were measured using Anton Paar MCR 702 rheometer. The viscosities were measured at the shear rates ranging from 10~$\mathrm{s}^{-1}$ to 100~$\mathrm{s}^{-1}$, within which all three experimental solutions were found to be Newtonian. The viscosity values of DI water, 10\% (v\slash v) glycerol and 3\% (w\slash v) ficoll 400, averaged over the shear rates were found to be 0.95~$\pm$~0.01, 1.32~$\pm$~0.02, and 1.48~$\pm$~0.01 mPa-s respectively.

To measure the activity of catalase in different media, the decomposition of its substrate, H$_2$O$_2$, was monitored using UV-Vis spectroscopy by recording its absorbance at 240~nm. Urease assay was performed by measuring absorbance of the dye phenol red at 560~nm. The enzyme and substrate concentrations were selected to ensure sufficient reaction times during tracer diffusion measurement. Student's t-test was performed to confirm the statistical significance of the experimental results. The measurements with a p-value less than 0.05 were considered to be statistically significant.

As mentioned earlier, the mechanical  fluctuations generated by active enzymes has previously been reported to be strong enough to influence the diffusion of nearby passive tracers under aqueous environments. A key question is whether such effects are also observable under cytosolic crowded conditions, where reactions catalysed by several enzymes lead to important physiological functions and life-sustaining processes. To check this, we performed experiments with molecules of ficoll 400 and glycerol to mimic cytocellular environments where the diffusion of 3~$\mu$m polystyrene (PS) tracers was studied in the presence of substrate catalysis by different enzymes. Concentrations of the crowders were chosen in such a way that their volume fractions were almost the same (10\%) with nearly identical viscosities (1.3-1.5 mPa-s). This is similar to the intracellular viscosity, although the cellular interior is highly inhomogeneous and compartmentalized that are often characterized by spatio-temporal viscosity variations.~\cite{42,43} Importantly, under crowded environment, the rate of substrate turnover by enzymes gets lowered owing to one or more of the following reasons: a) constrained diffusion of substrates to the enzymes' active sites, b) slower rate of conformation change of the enzymes during catalysis and c) accumulation of product molecules near the enzymes post catalysis~\cite{44,45,46}. In this scenario, it becomes crucial to explore whether the active fluctuations generated by enzymes can overcome dissipative effects and impact tracer diffusion in their vicinity. Our experimental results suggest that fluctuations generated by 1~nM catalase and 10~mM H$_2$O$_2$ were sufficient to influence the motion of the polymer microparticles in glycerol and ficoll. Similar observations were recorded with 10~nM urease and 100~mM urea which showed that this effect was likely to be generic and could very well be detectable for other enzyme-substrate pairs.

\begin{figure}[h]
    \centering
    \includegraphics[width=0.48\textwidth]{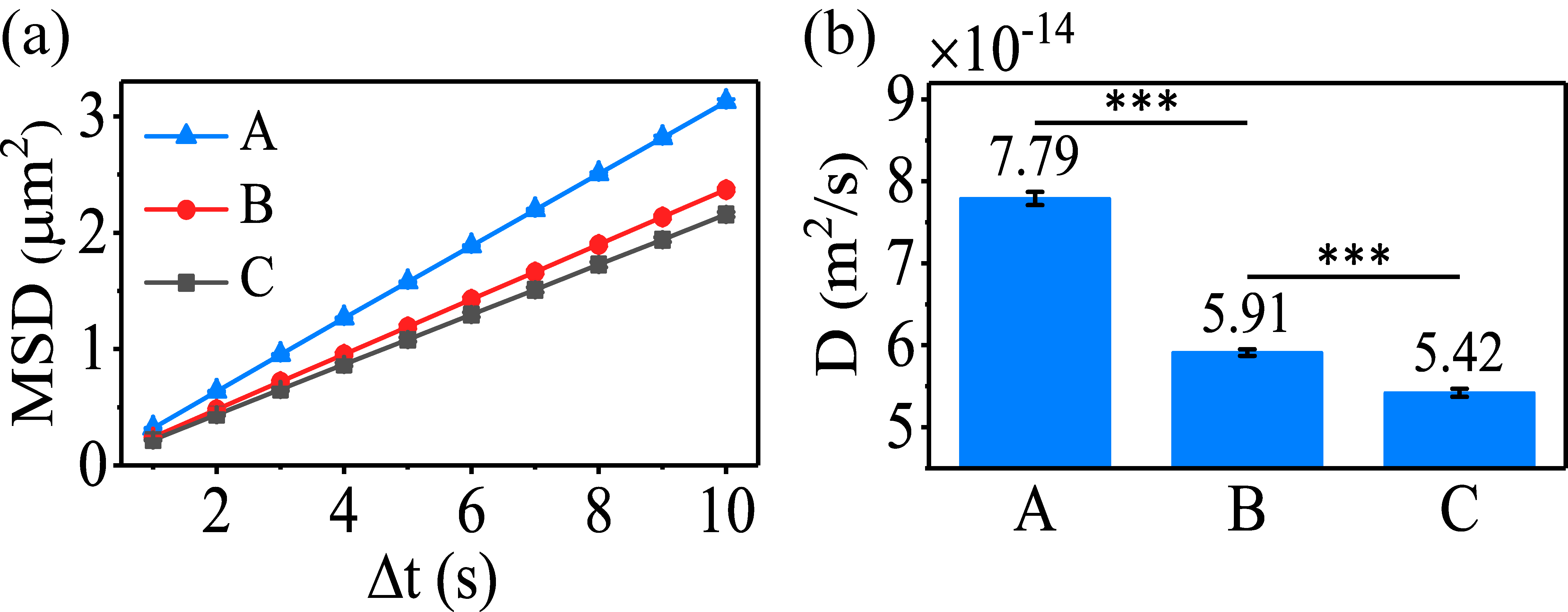}
    \caption{\justifying{(a) The mean-squared displacement (MSD) profiles, and (b) diffusion coefficients of 3~$\mu$m tracer particles in (A) DI water, (B) 10\% (v\slash v) glycerol, and (C) 3\% (w\slash v) ficoll 400. The symbol $\ast\!\ast\!\ast$ denotes the significance level of $p<0.001$.}}
    \label{fig:fig1}
\end{figure}

Initially, the tracer diffusion was measured in DI water where the diffusion coefficient was found to be 7.79 $\times$ $10^{-14}~\mathrm{m}^2/\mathrm{s}$, which was nearly 54\% less than the value predicted by the Stokes-Einstein equation (Fig.~\ref{fig:fig1}). This discrepancy may be attributed to the boundary effects on the particles which remained close to the bottom surface of the experimental chamber, at a height, h~$\sim$~5~$\mu$m. Faucheux and Libchaber characterized this hindered diffusion in terms of a dimensionless length scale parameter $\gamma$ given by, $\gamma = \frac{h-r}{r}$~\cite{47}. For our experiments with r = 1.5~$\mu$m particles, the value of $\gamma$ came out to be 2.33, which yielded a diffusion coefficient ratio $\frac{D}{D_0}$~$\sim$~0.55. This resulted in the estimated hindered diffusion coefficient to be D = 7.99 $\times$ $10^{-14}~\mathrm{m}^2/\mathrm{s}$, which matched nicely with our experimental results. The diffusion coefficients of the  polymer particles in glycerol and ficoll were measured to be 5.91 $\times$ $10^{-14}~\mathrm{m}^2/\mathrm{s}$ and 5.42 $\times$ $10^{-14}~\mathrm{m}^2/\mathrm{s}$ respectively, which also matched well with the estimated hindered diffusion coefficient values of 6.06 $\times$ $10^{-14}~\mathrm{m}^2/\mathrm{s}$ and 5.40 $\times$ $10^{-14}~\mathrm{m}^2/\mathrm{s}$  respectively (Fig.~\ref{fig:fig1}).

\begin{figure}[h]
    \centering
    \includegraphics[width=0.48\textwidth]{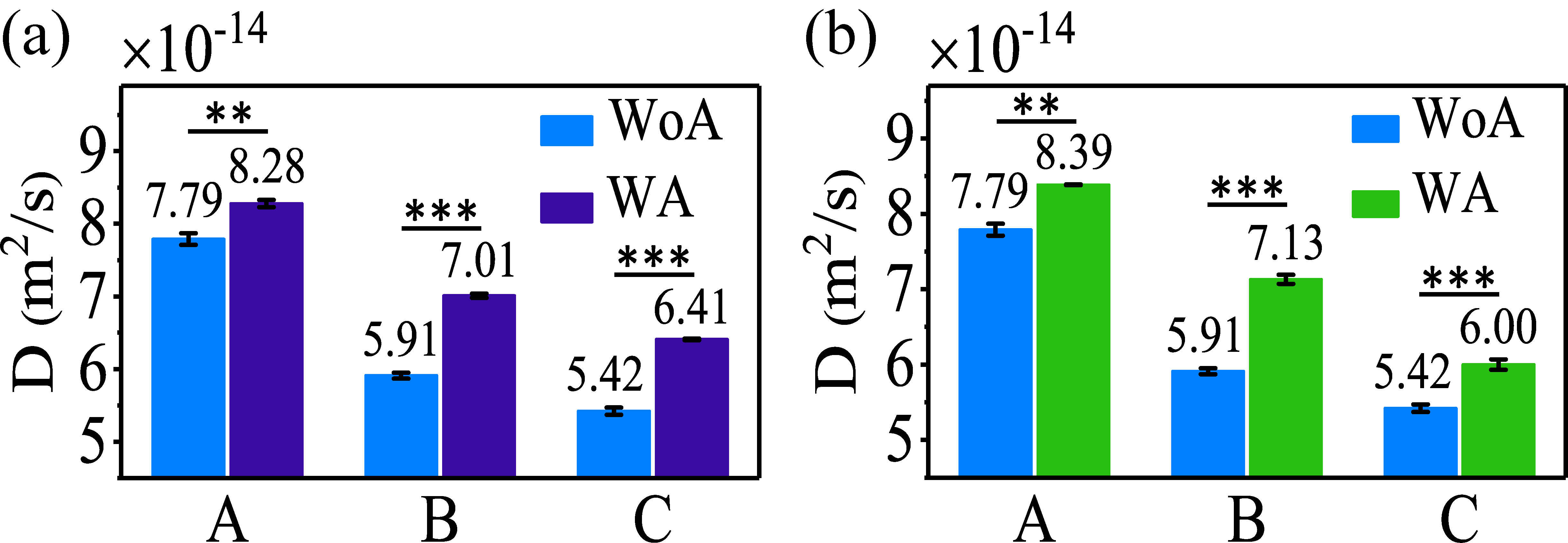}
    \caption{\justifying{Diffusion coefficients of tracer particles in the absence and presence of (a) catalase activity (1~nM catalase and 10~mM H$_2$O$_2$), and (b) urease activity (10~nM urease and 100~mM urea) in (A) DI water, (B) 10\% (v\slash v) glycerol, and (C) 3\% (w\slash v) ficoll 400 respectively. The symbols $\ast\ast$ and $\ast\!\ast\!\ast$ denote the significance levels of $p<0.01$ and $p<0.001$ respectively.}}
    \label{fig:fig2}
\end{figure}

We introduced non-thermal fluctuations in our system using two different biochemical reactions - the disproportionation of H$_2$O$_2$ with catalase and hydrolysis of urea with urease. Previous experiments conducted in aqueous environments demonstrated enhanced diffusion of passive tracers in the presence of free enzyme activity, where the increase in diffusion was found to correlate with enzymatic activity.~\cite{28} Importantly, consistent with  predictions of theoretical studies,~\cite{22,23,25} our experimental results presented herein confirmed that fluctuations generated by active enzymes are perceptible even in crowded environments and are capable of influencing the dynamics of their surroundings. 
Interestingly, we found that the relative enhancement in tracer diffusion is notably higher in crowded environments compared to that measured under aqueous conditions. In the presence of active catalse, the tracer diffusion was enhanced by 18.6\% and 18.3\% in 10\% glycerol and 3\% ficoll respectively (Fig.~\ref{fig:fig2}a). Similarly with active urease, diffusion enhancements of 20.6\% and 10.7\% were recorded in glycerol and ficoll respectively (Fig.~\ref{fig:fig2}b). 
To confirm that the free enzymes in solution did not get adsorbed over the polymer bead surface during experiments and influenced their propulsion, experiments were conducted with microparticles coated with a thin layer of bovine serum albumin (BSA). These coated microparticles also exhibited a similar enhancement in diffusion in the presence of catalysis. Furthermore, both BSA coated and uncoated tracers, when dispersed in fluorescently tagged enzyme solution showed no signs of enzyme attachment on their surfaces.

\begin{figure}[h]
    \centering
    \includegraphics[width=0.48\textwidth]{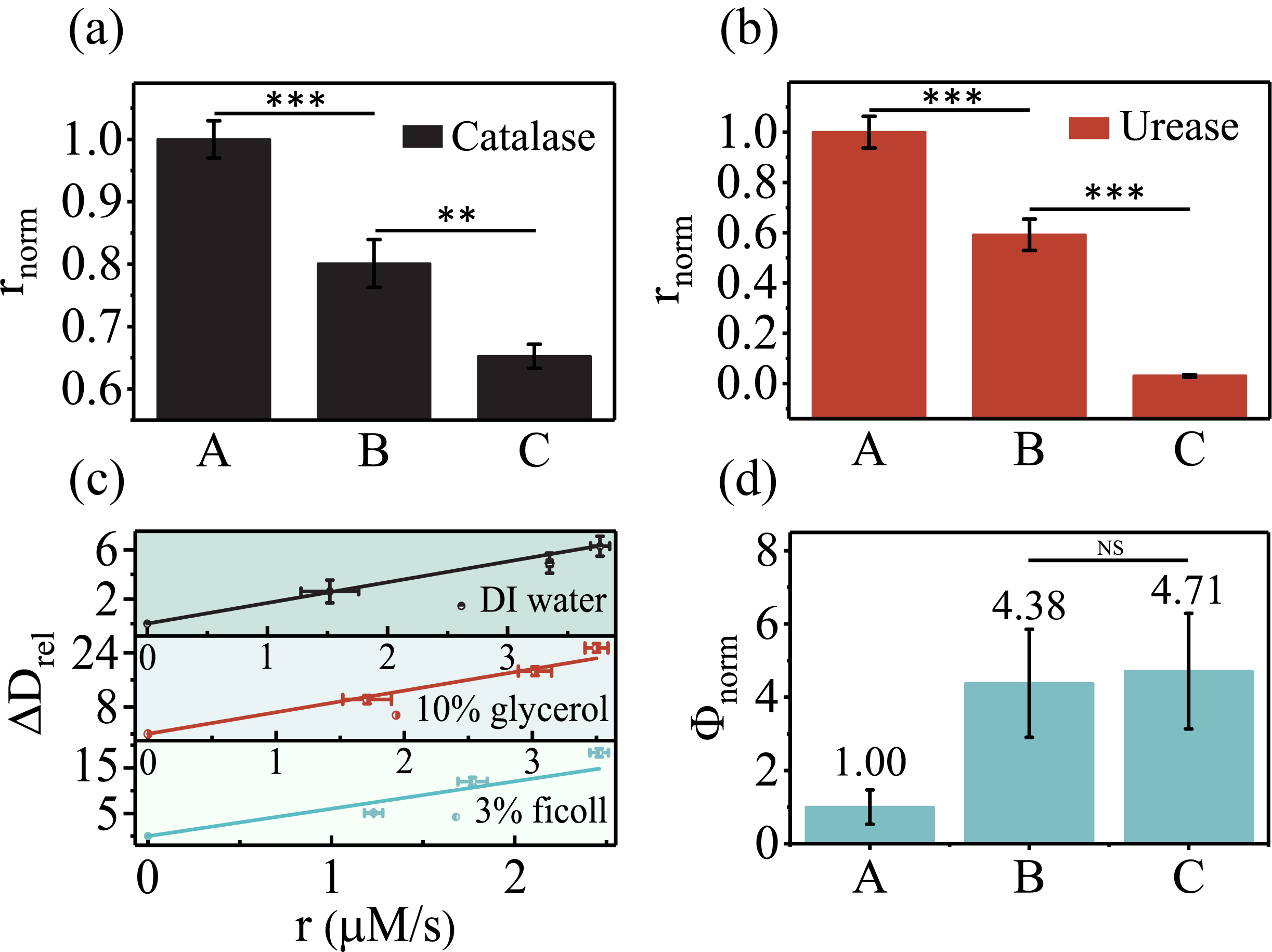}
    \caption{\justifying{Normalized reaction rates for (a) active catalase (1~nM catalase and 10~mM H$_2$O$_2$), and  (b) active urease (10~nM urease and 100~mM urea) in (A) DI water, (B) 10\% (v\slash v) glycerol, and (C) 3\% (w\slash v) ficoll 400. All the reaction rates were measured during 300-600 s from the beginning of the reactions. (c) Relative diffusion enhancement $\Delta$D$_{rel}$~of the tracer particles for different reaction rates r for active catalase in (A) DI water, (B) 10\% (v\slash v) glycerol, and (C) 3\% (w\slash v) ficoll 400. Experimental data were linearly fitted using OriginPro software. (d) Normalized relative tracer diffusion enhancement per unit reaction rate ($\Phi_{norm}$), estimated for different media. The symbols $\ast\ast$, and $\ast\!\ast\!\ast$ denote the significance levels of $p<0.05$ and $p<0.001$ respectively. NS signifies not significant.}}
    \label{fig:fig3}
\end{figure}

Considering the correlation observed between enzyme reaction rate and tracer diffusion in aqueous solutions, one possible explanation for greater enhancement in tracer diffusion within crowded environments could be the higher enzyme activity in crowded media, potentially boosting tracer diffusion therein. To assess the catalytic activity of catalase and urease in glycerol and ficoll, we conducted enzyme activity assays to determine the catalytic reaction rates r. We found that the substrate catalysis rates for catalase decreased by 19.9\% and 34.7\% in 10\% glycerol and 3\% ficoll respectively, compared to their rates in DI water (Fig.~\ref{fig:fig3}a). Similarly, for urease, the reaction rate decreased by 40.8\% and 96.9\% in glycerol and ficoll respectively (Fig.~\ref{fig:fig3}b). We further investigated the relative tracer diffusion, denoted as, $\Delta$D$_{rel}$ = $\frac{D-D_0}{D_0}~\times~100\%$ corresponding to different catalase reaction rates r, in different media (Fig.~\ref{fig:fig3}c). Assuming linearity between $\Delta$D$_{rel}$ and r, we calculated the rate of change of $\Delta$D$_{rel}$ with the reaction rate r from the slopes of (Fig.~\ref{fig:fig3}c). These slopes were subsequently normalized to obtain the relative enhancement in tracer diffusion in different crowded media compared to water, denoted as $\Phi_{norm}$, for identical reaction rate conditions. The variation of $\Phi_{norm}$ for different media is depicted in (Fig.~\ref{fig:fig3}d), which clearly indicates that the catalytic reaction rate was not primarily responsible for the higher energy transfer from enzymes to the tracers under artificial crowded environments.

The reaction generated heat can also be shown to be inefficient in boosting tracer diffusion in crowded solutions. In line with the estimations reported earlier,~\cite{40} simple calculations showed that upon each catalytic cycle, the increase in temperature across a single catalase molecule could amount to $\Delta$T~$\approx~10^{-5}$ K. Considering that the entire amount of heat generated upon substrate turnover was responsible for influencing nearby tracer diffusion, the corresponding thermophoretic tracer velocity could be estimated as $\mathrm{{V_{ST}}~\simeq~\frac{D_0 S_T \Delta T}{R}}$, where D$_0$, $\mathrm{S_T}$ and R are the normal diffusion coefficient, Soret coefficient and radius of the tracer particle. Considering $\mathrm{S_T~\sim~150~K^{-1}}$,~\cite{48} the change in the effective tracer diffusion coefficient due to thermophoresis is given by $\mathrm{\Delta D~\simeq~\frac{V_{ST}^{2}}{D_r}}$, where $\mathrm{D_r}$ is the rotational diffusion coefficient of the tracer. This yields, $\mathrm{\Delta D~\approx~10^{-20}~m^{2}/s}$ giving $\mathrm{\frac{\Delta D}{D_0}}~\approx~10^{-4}~\%$. This is too small to account for the experimentally observed enhancement in tracer diffusion upon catalysis. Moreover, considering that the crowder solutions used were sufficiently dilute, their thermal conductivities and resultant increase in tracer diffusion are expected to be similar to those of DI water.

Since neither the enzyme catalysis nor the reaction generated heat was found to be primarily responsible for greater enhancement in tracer diffusion in crowded environment, we next investigated the role of intervening media in facilitating energy transfer from the enzyme's active sites to the tracers. We developed a simple analytical model, where the binding and unbinding of substrates, resulting in conformational fluctuations in enzymes, were assumed to generate stochastic forces in the media. These forces were considered to be propagating around active enzymes in the form of high-frequency pressure pulses. For different actuation frequencies, we calculated the corresponding average energy densities at specific distances away from the point of actuation. The analysis finds its motivation from the fact that the intra-cellular environement is inherently viscoelastic, inhomegeneous and crowded, which is expected to play a key role in the propagation of enzyme-generated active fluctuations in its vicinity.
 
\section{\label{sec:level3}ANALYTICAL STUDIES}
\begin{figure}[h]
    \centering
    \includegraphics[width=0.48\textwidth]{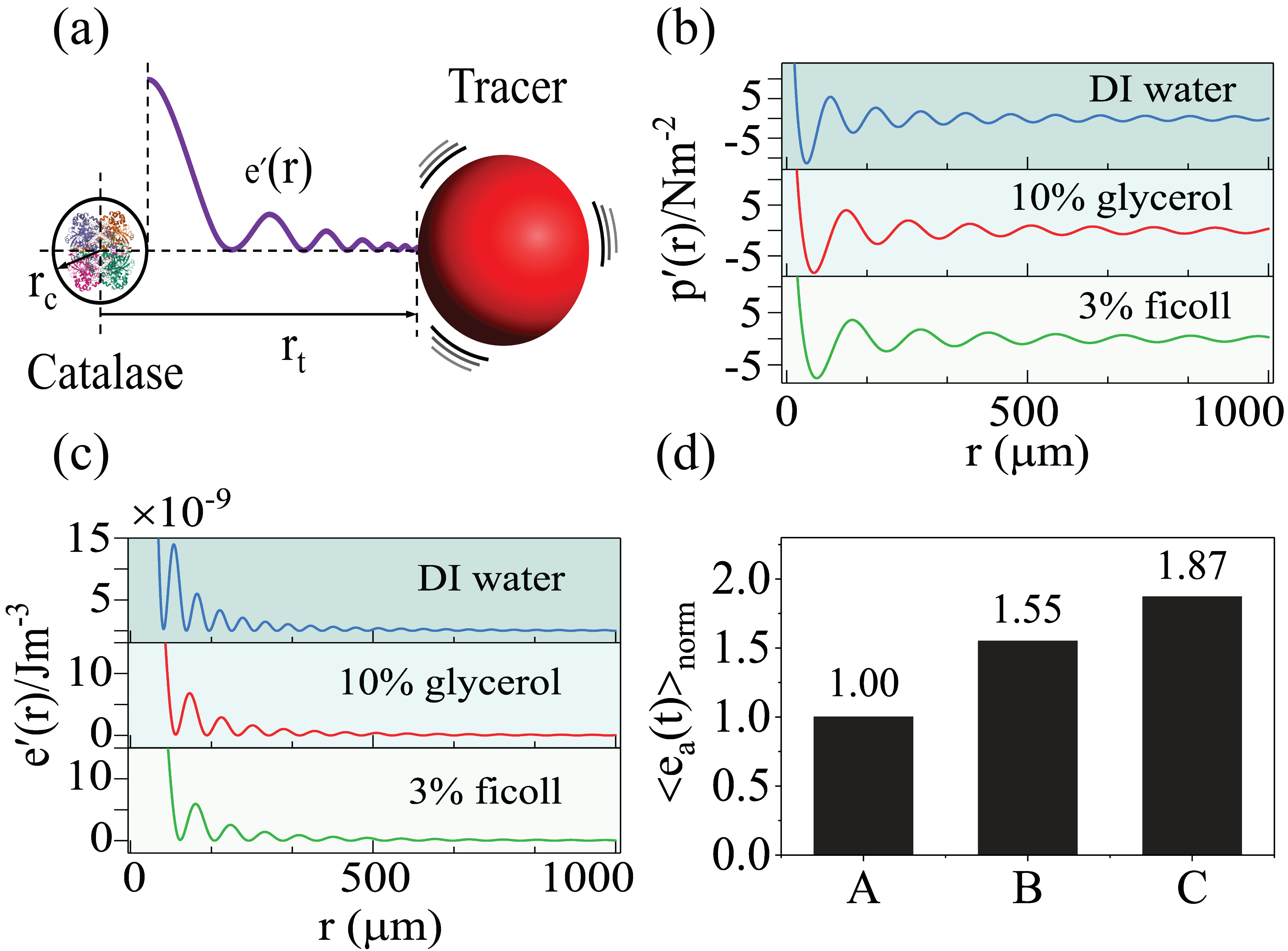}
    \caption{\justifying{(a) Schematic diagram of the enzyme catalase and a tracer particle. Spatial variation of the (b) pressure, and (c) the energy density at fixed instance of time in different medium. (d) Normalized root mean square value of the average energy density~($\langle e_a(t) \rangle_{norm}$) in (A) water, (B) 10\% glycerol, and (C) 3\% ficoll. Catalase structure images is taken from rcsb.org (PDB IDs: 1TGU).}}
    \label{fig:fig4}
\end{figure}

\protect\
We developed a qualitative understanding of the enhanced energy transfer in 10\% glycerol and 3\% ficoll, compared to that in water based on fundamental principles of fluid dynamics.~\cite{49} We considered the disturbances produced by catalytic reactions occurring at an enzyme site propagating through the media and reaching the tracer particles as spherical pressure waves. The energy density of these waves around an enzyme has been estimated. This is calculated by considering a simple case where an enzyme molecule was assumed to interact with single nearby tracer through reaction generated force fields, as shown in Fig.~\ref{fig:fig4}a. Starting from the equation of continuity, and using the Navier-Stokes equation for an isentropic compressible fluid with irrotational flows and negligible temperature gradient, we used the spherical polar coordinates to obtain the energy density as function of both space and time.

Although the timescale of conformational fluctuations corresponding to different domains of a single enzyme has been reported to vary over an extended range,~\cite{50} for our estimation, we assumed that the stress generated in the surrounding media following substrate binding and product release was related to one primary conformational change (which may or may not correspond to the catalytic step). We assumed the angular frequency of this primary conformational change to be $\omega$ in water, which generated an acoustic wave with a maximum pressure amplitude of $p_0$ at the surface of the enzyme. We also considered the pressure amplitude of this wave to be consistent in crowded conditions, although it propagated with a lower angular frequency in these media, primarily determined by the viscosity of the medium. This assumption found its rationale in previous experimental studies where the rate of conformation changes in enzymes was found to decrease with the increase in viscosity of the media $\eta$.~\cite{51} For our case, we took the angular frequency corresponding to the primary conformational fluctuation in crowded media to decrease as $\frac{1}{\eta}$. To estimate the wave number $k$ corresponding to the acoustic wave propagation in different media, we also considered their density $\rho_0$, velocity of sound $c_0$ therein, and viscosity $\mu_0$ values. The values considered and the corresponding angular frequencies ($\Omega$) for wave propagation in these media are given in Table.~\ref{property_table}.
\begin{table}[h!]
\begin{center}
\begin{tabular}{c|c|c|c|c}  
\hline
Medium & $\rho_0$ [kg/m$^3$] & $c_0$ [m/s] & $\mu_0$ [mPa-s] & $\Omega$ [s$^{-1}$]\\ 
\hline
Water & 1000 & 1481 & 1.00 & $\omega$\\
\hline
10\% glycerol & 1026 & 1532 & 1.32 & $\omega$/1.32\\
\hline
3\% ficoll & 1007 & 1504 & 1.48 & $\omega$/1.48\\ 
\hline
\end{tabular}
\caption{\justifying{Properties of different experimental solutions and their assumed effect on primary conformational fluctuation frequency of enzymes. The density and velocity of sound values for 10\% glycerol and 3\% ficoll are taken from~\cite{52,53} and~\cite{54}, respectively. The viscosity values of these solutions were measured experimentally.}}
\label{property_table}
\end{center}
\end{table}

The estimated wave number $k$ led to the calculation of enzyme fluctuation generated pressure $p'(r,t)$ and the velocity $u'(r,t)$ fields, along with the expression for the spatiotemporal variation of energy density $e'(r,t)$. 
\begin{eqnarray}
    \label{psol_spec} 
    p'(r,t) = \text{Re}\left[\frac{p_{0}r_c}{r} e^{i\{k(r-r_c)-\Omega t\}}\right]\\
    \label{usol_spec} 
    u'(r,t) = \text{Re}\left[\frac{p_0 \Omega r_c}{\rho_0 c_0^2} \left( \frac{1}{kr} + \frac{i}{k^2r^2} \right) e^{i\{k(r-r_c)-\Omega t\}}\right]\\
    \label{esol_spec} 
    e'(r,t) = \frac{\rho_0}{2} u'(r,t)^2 + \frac{1}{2 \rho_0 c_0^2} p'(r,t)^2
\end{eqnarray} 
Here, $r_c$ is the radius of the enzyme. We also considered $r_t$ to be the distance between the center of the enzyme and the tracer surface (Fig.~\ref{fig:fig4}a). The average energy density over a volume bounded by $[r_c, r_t]$ is then given by
\begin{eqnarray}
    \label{eavg_spec} 
    e_a(t) = \frac{4\pi}{V_0} \int_{r_c}^{r_t} e'(r,t) r^2 dr
\end{eqnarray} 
where, $V_0=4\pi (r_t^3 -r_c^3) /3$ is the volume of the spherical shell with its center coinciding with the center of the enzyme (Fig.~\ref{fig:fig4}a). 

The integral in Eq.~\eqref{eavg_spec} was evaluated numerically using the Gauss-Legendre quadrature method. For quantitative estimates, we considered $\omega=10^8$ Hz and $r_t = 1000~\mu$m. Considering the force generated per catalytic turnover $\sim$~$10$ pN, as reported earlier~\cite{8,55} and enzyme radius $r_c\sim$~5 nm, the instantaneous pressure generated per catalytic turnover came out to be $\sim$~0.1 MPa. We took the value of maximum pressure amplitude $p_0$ to be of this order, which resulted in the spatial variation of the pressure and energy density at a fixed instance of time, as shown in Fig.~\ref{fig:fig4}b and~\ref{fig:fig4}c, respectively. We then obtained the root mean square value of the average energy density, $\langle e_a(t) \rangle$, over a time period corresponding to the actuation frequency. The ratio of $\langle e_a(t) \rangle$ for glycerol and ficoll with respect to water, represented as~$\langle e_a(t) \rangle_{norm}$, has been shown in Fig.~\ref{fig:fig4}d. The estimated trend of~$\langle e_a(t) \rangle$$_{norm}$ variation qualitatively matched well with the experimentally measured energy transfer, characterized by $\Phi_{norm}$, as shown in Fig.~\ref{fig:fig3}d. We performed the analysis for two other values of ~$\Omega$ ($10^{7}$ and $10^{9}$ Hz), which also resulted in similar trend of~$\langle e_a(t) \rangle$$_{norm}$ in different experimental media. We finally noted that the estimated energy transfer values (measured by~$\langle e_a(t) \rangle$$_{norm}$) were relatively lesser in magnitude compared to the experimentally measured values of $\Phi_{norm}$. The difference in the magnitudes may be attributed to the assumptions made in the model - such as the consideration of (1) interaction between a single enzyme and tracer and disregard for potential influences from multiple particles, (2) only a single frequency of conformational fluctuations responsible for energy transfer from the enzyme to the tracer. It may be noted that the viscosity of the medium played a key role only in reducing the actuation frequency of the enzyme as mentioned earlier. Owing to the small length scales of the system, the attenuation of the pressure wave had negligible dependence on the viscosity of the medium. The sharp decrement in the spatial components of $p'(r,t)$, $u'(r,t)$, and $e'(r,t)$ with distance were solely due to the spherical nature of the reaction generated pressure waves.

\section{\label{sec:level4}CONCLUSIONS}\protect\
Our experimental findings suggested that the change in tracer diffusion ($\Delta$D) per unit enzyme reaction rate for 10\% (v\slash v) glycerol and 3\% (w\slash v) ficoll 400 are 4.38 and 4.71 times higher than that measured in aqueous environment (Fig.~\ref{fig:fig3}d). Analytical estimates based on the theory of wave propagation in fluids also suggested that the energy transfer from active enzymes to their immediate neighbourhood enhanced in glycerol and ficoll compared to that in aqueous media (Fig.~\ref{fig:fig4}d). Thus our observation led to the insights into the propagation of enzymatic forces in active crowded environments. The augmented transport of molecules and particles through enzyme generated non-thermal mechanical fluctuations carries significant scientific and technological implications. The generation of forces induced by catalysis may provide an adequate explanation for the observed stochastic motion of the cytoplasm, cytoplasmic glass transitions, and improved mixing observed during metabolic transformations in bacterial cells.~\cite{33,56} The movement of proteins within the cytoplasmic matrix is vital for their effective function across various organelles, and enzyme-generated forces could potentially serve as key orchestrators of this intricate dynamic interplay.

\begin{acknowledgments}

KKD thanks Science and Engineering Research Board (SERB), India (ECR/2017/002649, CRG/2023/007588), Department of Science and Technology (DST), India (DST/ICD/BRICS/PilotCall3/BioTheraBubble/2019), Ministry of Education, Government of India (MoE-STARS/STARS-2/2023-0620) and IIT Gandhinagar for financial support. We also thank Sanchari Swarupa, Aditi Saha, Ayushi Bhatt, Sirsha Ganguly, and Nividha for insightful discussions. RC thanks IIT Gandhinagar for a research fellowship. 
\end{acknowledgments}

\bibliography{manuscript}
\end{document}